\documentclass[prl,aps,a4,epsf,8pt,twocolumn,showpacs]{revtex4}

\usepackage{amsmath}
\usepackage{amssymb}
\usepackage{graphicx}
\usepackage[usenames]{color}

\begin{document}

\newcommand{\beq}{\begin{equation}}
\newcommand{\eeq}{\end{equation}}
\newcommand{\beqa}{\begin{eqnarray}}
\newcommand{\eeqa}{\end{eqnarray}}

\newcommand{\eq}[1]{Eq.~(\ref{#1})}

\title{Vortex solid phase with frozen undulations in superconducting
   Josephson-junction arrays in external magnetic fields}

\author{Hajime Yoshino$^{1}$, Tomoaki Nogawa$^{2}$, Bongsoo Kim$^{3}$
}
\address{$^1$Department of Earth and Space Science, Faculty of Science,
 Osaka University, Toyonaka 560-0043, Japan\\
$^2$Department of Applied Physics. School of Engineering, 
The University of Tokyo, 7-3-1 Hongo, Bunkyo-ku, Tokyo 113-8656, Japan\\
$^3$Department of Physics, Changwon National University, 
Changwon 641-773, Korea.
}

\begin{abstract}
A vortex solid with self-generated randomness is found theoretically in
 a frustrated Josephson junction array (JJA) under external magnetic
 field with anisotropic couplings. Vorticies induced by external
 magnetic field develop stripes parallel to the direction of weaker
 coupling. It is shown analytically that there is a continuous,
 gapless band of metastable states in which stripes are deformed
 randomly by transverse undulation. The vortex solid with the frozen
 undulation in a metastable state freely slides along the direction of
 stronger coupling, thereby destroying ordering of phases even at zero temperature, but is jammed along the direction of weaker coupling.
\end{abstract}


\date{\today}
\maketitle

\begin{figure}[t]
\includegraphics[width=0.45\textwidth]{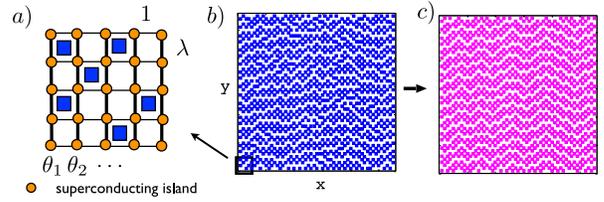}
\caption{Vortex patterns in an irrationally frustrated JJA under
external magnetic field with anisotropic coupling. Here
 $\lambda=1.5$ so that the coupling is  stronger along $y$ direction. 
Such an anisotropic JJA can be fabricated in laboratory 
by lithography technique(s) \cite{anisotropic-JJA}. 
Panel a) displays the JJA on a square lattice.  A fraction 
 $f=21/55$, which approximates an irrational number
 $(3-\sqrt{5})/2=0.381966..$, of the plaquettes are occupied by vorticies
with charge $1-f$ represented by filled squares. 
Panel b) displays an equilibrium vortex pattern at $T=0.2$
 and Panel c) displays that at a nearby energy minimum reached via an
 energy  descent algorithm.  
}
\label{fig-stripe}
\end{figure} 

Solids are systems with rigidity ranging from crystals, quasi-crystals to
glasses and jammed granular matters \cite{Nagel-group}.  
Generally it becomes more challenging to understand mechanism of the
formations of {\it less} periodic solids distinguishing them from liquids \cite{Kurchan-Levine}. 
A useful guiding concept to study non-crystalline solids is 
{\it frustration} which inhibits simple crystallization. 
Imagine that there is a trick to ``inject'' dislocations artificially into a crystal
from outside and that their density can be controlled at will. 
Such a system will provide a very interesting ground to study
consequences of geometrical frustration, especially realization of {\it self-generated randomness} or {\it glassiness without quenched disorder} \cite{Sadoc,Tarjus-review}.
Quite remarkably the Josephson
junction array (JJA) under external magnetic field realizes such an ideal
situation \cite{Tinkam,JJA-exp-reviews,Teitel-Jayaprakash-1,Halsey,anisotropic-JJA}. Furthermore, transport properties of the JJA under external current can be regarded as ``rheology'' under external shear \cite{vortex-jamming,from-FK-to-JJA}. 

JJA  is a regular network of superconducting islands connected with each other by
Josephson junction in the form, say, of a square lattice of size $N=L
\times L$ as shown in Fig.~\ref{fig-stripe} a)  \cite{JJA-exp-reviews,Tinkam}. The phases $\theta_{i}$
of the superconducting order parameter on the islands $i=1,2,\ldots,N$
interact with each other via Josephson coupling. 
Under magnetic field $B$, which can be varied at will in experiments,
the number density $f = B a^{2}/\phi_{0}$ of vorticies (dislocations) can
be forced into the configuration of the phases. Here $a^{2}$ is the area
of a plaquette  and $\phi_{0}$ is the flux quantum. 

An interesting connection of JJA to the problem of {\it frictions} provides
valuable insights. The frustrated JJA becomes essentially equivalent to
the Frenkel-Kontorova (FK) model \cite{FK-review} and 
the two-chain model of  Matsukawa and Fukuyama (MF)
\cite{matsukawa-fukuyama} in one dimensional limit ($N=L \times 2$), i. ~e. 
on the ladder lattice \cite{from-FK-to-JJA}\cite{ladder-JJA}. 
These one-dimensional systems are known to undergo a kind of jamming or frictional transition 
at zero temperature $T=0$, known as the Aubry's transition
\cite{aubry,kawaguchi-matsukawa} at a critical value of the
strength of coupling $\lambda$ between two surfaces which are {\it
incommensurate} with respect to each other. Then one would naturally be
led to consider {\it irrationally frustrated anisotropic} JJA
\cite{vortex-jamming} with 1) {\it irrational} vortex density $f$
\cite{Teitel-Jayaprakash-1,Halsey} and 2) {\it anisotropic} 
couplings into $x$ and $y$ directions, say $1$ into $x$ direction 
and $\lambda$ into $y$ direction.

In this Letter we study the ground state as well as low-lying states of the
irrationally frustrated JJA with sufficiently strong anisotropy $\lambda \gg 1$. 
Numerically we found vortex stripes parallel to the
direction of weaker coupling. 
In addition to the ground state in which vortex stripes are straight,
we found numerous metastable states with different
realizations of transverse undulation of the stripes 
as shown in Fig.~\ref{fig-stripe}.
By a perturbative analysis in series of $1/\lambda$ starting from 
infinite anisotropy limit $\lambda=\infty$, 
we are able to reproduce a family of such low-lying metastable states including the ground state analytically.
The coexistence of sliding and jamming in the system
\cite{vortex-jamming} is proved from the analytically constructed ground state and the low-lying states.
Because of the sliding, the phases remain disordered even at $T=0$
for irrational $f$, in sharp contrast to JJA with rational $f$ where not only vorticies 
(chiralities for $f=1/2$) but also phases exhibit (quasi-)long ranged order at $T>0$ 
\cite{Teitel-Jayaprakash-2}.

{\bf Model} To simplify notations we label the vertices
(superconducting islands) as $i=1,2,\ldots,N$ 
whose position in the real space is given by $(n_{i},m_{i})$. 
The properties of the JJA under the transverse magnetic field
are known to be described by an effective classical Hamiltonian \cite{Tinkam},
\beq
H= -\sum_{<i,j> \parallel x\mbox{-axis}} \cos(\psi_{ij})
 - \lambda \sum_{<i,j> \parallel y\mbox{-axis}} \cos(\psi_{ij})
\label{eq-jja}
\eeq
where $<i,j>$ denotes nearest neighbor and $\psi_{ij}$
the gauge-invariant phase difference, $\psi_{ij} \equiv \theta_{i}-\theta_{j}-A_{ij}$.
The temperature $T$ is defined in a unit with $k_{\rm B}=1$.
For the anisotropy $\lambda$, we need to consider only $\lambda > 1$ 
by symmetry. The vector potential $A_{ij}(=-A_{ji})$ is defined such that directed
sum of them around each plaquette is  $2\pi f$.

Vortex charge $v_{i}$ of the vortex  at the plaquette associated with the $i$-th vertex is defined by
taking directed sum of $(\psi_{ij}/2\pi-[\psi_{ij}/2\pi]_{n})$
on the junctions around the plaquette. Here $[x]_{n}$ denotes the nearest integer of the real variable $x$. It takes values $\ldots,-1-f,-f,1-f,\ldots$. 
We use periodic boundary conditions so that the total vortex charge is enforced to be zero (charge neutrality).

It has been proposed that {\it superconducting glass} may be realized if $f$ is
{\it irrational} \cite{Halsey}. While JJAs with rational $f$
develop periodic vortex lattices
\cite{Teitel-Jayaprakash-2,ground-states-isotropic}, 
such simple orderings may be avoided with
irrational $f$. Indeed equilibrium relaxations were similar to the
primary relaxations observed in typical fragile supercooled liquids
\cite{JJA-relaxation}. Such a system is called as {\it irrationally
frustrated} JJA. \cite{comment}

{\bf Numerical methods} 
In numerical simulations, we used a series of rational numbers 
$p/q=5/13$, $8/21$, $13/34$, $21/55$, $34/89$, $55/144$, $89/233$ for the filling $f$, which approximate an irrational number $f=(3-\sqrt{5})/2=0.38196601...$
Square lattices of size $L \times L$ with periodic boundary
conditions for both directions are used. We choose
$L=q$ so that the ratio $f=p/q$ converges to the target irrational number
in $L \to \infty$ limit. 

To generate the equilibrium ensemble, we used 
exchange Monte Carlo (MC) simulations combined with the over-relaxation method
\cite{method} performed on systems with $L=13-89$ using $20-120$ temperatures 
in the temperature range $T=0.2-0.4$. 
We used $10^{5}-10^{6}$ MC steps for the equilibration and observations.

\begin{figure}[t]
\includegraphics[width=0.45\textwidth]{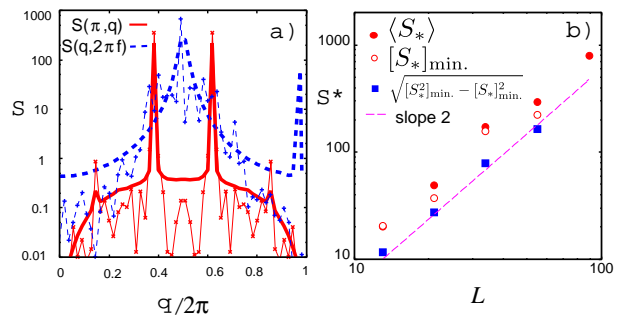}
\caption{Structure factor of vorticies.
a) displays the cross-sections of the structure factor $S(q_{x},q_{y})$ ($L=55$) with 
thermal average (thick lines) and at the energy minimum shown in Fig.~\ref{fig-stripe} c)
(thin lines).
b) displays the amplitude of the peak of the structure factor
 $S_{*}=S(\pi,2\pi f)$ with thermal average $\langle S_{*}\rangle$,
 average over the energy minima $[S_{*}]_{\rm min.}$ and variance of the
 minima-to-minima fluctuation $\sqrt{[S_{*}^{2}]_{\rm min.}-[S_{*}]_{\rm
 min.}^{2}}$. Here the average over minima $[\ldots]_{\rm min}$ is taken over 
$100$ energy-minima obtained by independent initial conditions.} 
\label{fig-structure-factor}
\end{figure}

{\bf Stacked undulating vortex stripes}
As shown in Fig.\ref{fig-stripe}, 
the vorticies develop undulated stripes parallel to
the direction of weaker coupling at low temperatures. The formation of the 
stripes is reasonable because the repulsive interactions between
vorticies are anisotropic. A remarkable feature is that the stripes
are stacked quite regularly along the stronger coupling 
as shown in Fig.\ref{fig-stripe} c)  
in the nearby energy minima obtained via a simple energy descent
algorithm. Starting from different thermalized configurations we obtained
numerous energy minima similar to the one shown in Fig.\ref{fig-stripe}
c) but with different realizations of the transverse undulation. 
The nearly perfect stacking of the stripes strongly suggests
that the energy barrier in going from one to another realization of significantly different 
undulation of vorticies (dislocations), which necessarily involve large number of
plastic deformations, diverges with the system size so that the ergordicity is broken. 
This feature is markedly different from
usual undulations found, for example, in liquid
crystals which are fluctuating dynamically \cite{Chaikin-Lubensky}.

The stacked undulation is manifested in the structure factor of the vorticies.
As shown in  Fig. \ref{fig-structure-factor}, the structure factor
$S(q_{x},q_{y})$ exhibits prominent peaks at $(q_{x},q_{y})=(\pi,2\pi
f)$ and $(\pi,2\pi(1-f))$ whose height scales with the system size as
$N=L^{2}$ as usual Bragg peaks do. However the profile of the peak is
peculiar: it decays sharply along $q_{y}$ reflecting the stacking 
but decays slowly by a power law $|\pi-q_{x}|^{-2}$ along $q_{x}$
reflecting the transverse undulation.  Here we emphasize again that the transverse
undulation is frozen in time. 
The frozen-in randomness is manifested in the minimum-to-minimum
fluctuations of the structure factor shown in
Fig.\ref{fig-structure-factor} a) and b). Note that the variance of the
fluctuation as well as the average grows linearly with the system size
$N$ meaning that the structure factor is {\it not} self-averaging.

{\bf Analytic construction of the ground state}
Let us now turn to explicit construction of low-lying states by an
analytical approach. To this end we propose a non-trivial ansatz for the ground state using the notion of the so called hull functions developed in the studies of the FK and MF models  \cite{aubry,ladder-JJA,matsukawa-fukuyama,kawaguchi-matsukawa,FK-review}.
We propose that the gauge invariant phase differences $\psi_{ij}$ across the Josephson junctions, 
in the low-lying states of the anisotropic JJA ($\lambda > 1$) can be represented  as,
\beq
\psi_{(x,y) (x+1,y)}= \phi_{x}[y+\alpha(x)]  \qquad
 \psi_{(x,y) (x,y+1)}=\phi_{y}[y+\alpha(x)]
\eeq
Here $\phi_{x}[y]$ and $\phi_{y}[y]$ are functions defined on the
``folded coordinate'' $[y]=fy-{\rm int}(fy)$ where ${\rm int (x)}$
is the floor function. The folded coordinate takes values limited in
the range $0  \leq  [y] < 1$. Such a function is called as a hull function \cite{aubry,FK-review}.
Note that if $f$ is {\it irrational}, which we always assume in this
work, the vertices of the JJA
{\it uniformly} fill the entire range of the folded coordinate $[y]$ in
the limit $N \to \infty$.
Thus we can treat $[y]$ as a {\it continuous} variable.
Moreover one can then easily extract distribution of the values of the phase
differences from the hull function because
of the uniform distribution of $[y]$ over the support $0 \leq   [y] < 1$.

An obvious constraint 
is that the directed sum over $\psi_{ij}$ around each plaquette 
must be $-2\pi f$.
In addition, the Josephson currents must be conserved  at each
vertex (force balance condition) in each energy minimum. Thus 
the following two conditions should hold,
\beqa
&&\hspace*{-0.5cm} \phi_{x}[y]+\phi_{y}[y+\delta(x)]-\phi_{x}[y+1]-\phi_{y}[y]=-2 \pi f
\label{eq-condition-plaquette} \\
&&\hspace*{-0.5cm}\frac{1}{\lambda}\sin \phi_{x}[y]+\sin \phi_{y}[y] 
= \frac{1}{\lambda}\sin \phi_{x}[y-\delta(x)]+\sin \phi_{y}[y-1]  \hspace*{.5cm}
\label{eq-condition-vertex}
\eeqa
where $\delta(x)\equiv\alpha(x)-\alpha(x-1)$.

Now our task is to look for the hull functions $\phi_{x}[y]$,
$\phi_{y}[y]$ and phase differences $\delta(x)$ which satisfy 
the conditions on the plaquettes \eq{eq-condition-plaquette}
and vertexes \eq{eq-condition-vertex}. We solve this problem by
performing a $1/\lambda$ expansion 
\cite{kawaguchi-matsukawa,Nogawa-Nemoto}around the infinite anisotropy limit $\lambda=\infty$.

In the infinite anisotropy limit $\lambda \to \infty$ 
the weaker couplings can be neglected so that we easily find
$
\phi_{x}[y]=(2 [y]-1) \pi +O(1/\lambda)$ and $\phi_{y}[y]=O(1/\lambda)$ 
which trivially satisfies \eq{eq-condition-plaquette} and
\eq{eq-condition-vertex} (with $\lambda=\infty$). 
As such, the phase difference $\delta(x)$ is not fixed at this stage.

Let us consider first the ground state assuming
that phase difference is uniform, ~i.~e. $\delta(x)=\delta$.
Using the above results in \eq{eq-condition-vertex} we find
$1/\lambda$ correction term of $\phi_{y}[y]$, which in turn allows us
to find $1/\lambda$ correction term of $\phi_{x}[y]$ through
\eq{eq-condition-plaquette}. In this way we obtained analytic form of the
hull functions up to $O(\lambda^{-3})$ as,
$\phi_{x}[y]=(2 f y - 1)\pi+\frac{|a_{1}(\delta)|^{2}}{\lambda} s_{x}(1,y)
+\frac{|a_{1}(\delta)|^{4}}{8\lambda^{3}} 
\left ( |a_{3}(\delta)|^{2}  s_{x}(3,y)
-3 |a_{1}(\delta)|^{2} s_{x}(1,y) \right)$ and
$\phi_{y}[y]=\frac{|a_{1}(\delta)|}{\lambda}s_{y}(1,y) 
+\frac{|a_{1}(\delta)|^{4}}{8\lambda^{3}} 
\left (|a_{3}(\delta)| s_{y}(3,y) 
 - 3 |a_{1}(\delta)| s_{y}(1,y) \right)$
with $a_{n}(\delta) \equiv (1-e^{-i 2\pi f \delta})/(1-e^{-i2 n \pi f})$,
$s_{x}(n,y) \equiv \sin(2n\pi f y)$ and 
$s_{y}(n,y)\equiv \sin(2n\pi f y+{\rm Arg}\;a_{n}(\delta))$. 
For {\it irrational} $f$, we find the energy is minimized 
by choosing $\delta=\delta^{*}\equiv 1/(2f)$ and obtain the ground state 
energy $E_{\rm g}$, 
$\frac{E_{\rm g}/\lambda}{N}=-1-\frac{|a_{1}(\delta^{*})|^{2}}{4
\lambda^{2}}-\frac{|a_{1}(\delta^{*})|^{4}}{16\lambda^{4}}\left (\frac{1}{4}-|a_{1}(\delta^{*})|^{2}
\right)+O(\lambda^{-6})$.
The vortex configuration of the ground state is indeed the stripes like
Fig.~\ref{fig-stripe} but without the transverse undulation.

{\bf Band of undulated metastable states} Next let us construct the low-lying
states with transverse undulation of stripes shown in Fig.~\ref{fig-stripe}.
Somewhat surprisingly, we can solve 
\eq{eq-condition-plaquette} and \eq{eq-condition-vertex} 
with {\it arbitrary} $\delta(x)$, finding,
$\phi_{x}[y] =(2 f y - 1)\pi+
-\frac{|a_{1}(\delta^{*})|^{2}}{4\lambda}\left[ C_{1}[\delta(x)]\cos(2\pi f y) +C_{2}[\delta(x)]s_{x}(1,y))\right] 
 +O(\lambda^{-2})$ and 
$\phi_{y}[y] = \frac{|a_{1}(\delta^{*})|}{2\lambda}
 \left( s_{y}(1,y-\delta(x)+1/(2f)) +s_{y}(1,y) \right)+O(\lambda^{-2}) \nonumber$
with
$C_{1}[\delta(x)] \equiv -\sin(2\pi f \delta(x+1))-\sin(2\pi f \delta(x))$
and
$C_{2}[\delta(x)] \equiv -2+\cos(2\pi f \delta(x+1))+\cos(2\pi f \delta(x))$.

As the result the energy becomes, for {\it irrational} $f$,
$\frac{E_{\rm g}}{\lambda}
+L\frac{|a_{1}(\delta^{*})|^{2}}{8 \lambda^{2}}
\sum_{x=1}^{L}(1+\cos (2\pi f \delta(x)))+O(\lambda^{-4})$. 
It is evident that there exists a gapless, continuous spectrum of low-lying states
each of which is parametrized by a function $\delta(x)$. 
Assuming $|\delta (x)-1/(2f)| \ll 1$,  we
obtain an one dimensional {\it elastic} Hamiltonian  with an
unusual elastic constant which grows linearly with the system size $L$.

Let us emphasize that the undulated states with the arbitrary
$\delta(x)$ are ensured to {\it satisfy the force balance
condition} \eq{eq-condition-vertex}. 
Thus the system trapped in an undulated 
state {\it cannot} relax {\it spontaneously} down to the ground state
since they are {\it metastable}: the undulation is distinct from ``phonons'' by which vorticies cannot move.

\begin{figure}[t]
\includegraphics[width=0.45\textwidth]{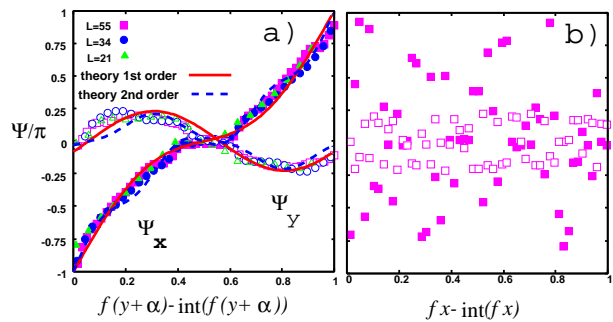}
\caption{Configuration of the gauge invariant phase differences
across Josephson junctions in energy minima.
Here the anisotropy is $\lambda=1.5$.
In panel a) original $y$ axis is ``folded'' to $[y]=fy-{\rm int}(fy)$.
The symbols are data of the gauge invariant phase differences 
$\psi_{x}=\psi_{(x,y) (x+1,y)}$ (filled symbols)
 and $\psi_{y}=\psi_{(x,y) (x,y+1)}$ (open symbols) across
Josephson junctions parallel to $x$ and $y$ axis.
Each data set consists of data points on a 'column'
 $(x,1),(x,2),\ldots,(x,L)$ at an arbitrary chosen $x$ ($1 \leq x \leq L$) 
in an arbitrary chosen energy minima respectively. 
Each data set is shifted globally by 
some $\alpha$ so that different data sets collapse on top of each other.
The lines are analytically
 obtained hull functions of the ground state. In panel b)
 the data on the same energy minimum ($L=55$) are plotted  against ``folded'' $x$-axis.
}
\label{fig-hull}
\end{figure} 

Now the unusual structure factor of the vorticies $S(q_{x},q_{y})$ in 
Fig.~\ref{fig-structure-factor} can be understood as follows. 
The configuration of the ground state is a function of the folded coordinate $[y]$ 
so that $S(q_{x},q_{y})$ must 
have peaks along the $q_{y}$-axis at $q_{y}=2\pi f$  and $2\pi (1-f)$. 
On the other hand the phase shift of the hull function 
$\delta^*=1/(2f)$ along $x$-direction meaning that $S(q_{x},q_{y)}$ must
have a single peak along the $q_{x}$ axis at $q_{x}=\pi$, reflecting the
horizontal stripes. 
The power law tail $|q_{x}-\pi|^{-2}$ naturally follows from 
the effective one-dimensional elastic Hamiltonian for the transverse
undulation obtained above.

In Fig.~\ref{fig-hull} we plot the phase differences across the junctions
in energy minima obtained numerically and compare them with the hull
functions obtained analytically. For simplicity we show here the hull function for
the ground state disregarding small differences due to the undulation. 
Our perturbative result grasps well the overall features. 
The agreement will be further improved by taking into account higher
order terms in the $1/\lambda$ expansion.

A remarkable consequence of the analytic hull function
is that the undulated vortex solid can {\it freely
slide} into the $y$ direction: 
Given an energy minimum described by the hull functions $\phi_{x}[y]$ and
$\phi_{y}[y]$, a family of solutions with exactly the same energy can be 
obtained through the operation $[y]\to [y+\alpha]$ with varying phase 
shift parameter $\alpha$. 
Consequently the phases must remain disordered even at $T=0$.
As shown in Fig.~\ref{fig-hull} a) the phase differences across
junctions 
parallel to the $x$-axis take all possible values in the range 
$-\pi \leq \phi_{x}[y] < \pi$ meaning that the system can be {\it sheared}
indefinitely along the $x$-axis
which amount to unidirectional sliding of the undulated vortex solid into the
$y$-direction with fixed pattern. In contrast, the plot against ``folded''
$x$-axis shown in Fig. ~\ref{fig-hull} b) exhibits no hint of a single
valued, regular hull function. 
Also note that the distribution of the
$\phi_{y}$ does not span the entire range
$-\pi  \leq \phi_{y} < \pi$ which is needed to allow shear along
$y$-axis measning jamming along $x$-axis.

It is instructive to compare the above results with the FK model.
In the FK model the hull function is proved to be an 
analytic function in the sliding phase $\lambda
<\lambda_{c}$ but becomes non-analytic in the jammed phase $\lambda >
\lambda_{c}$ \cite{aubry}. In the anisotropic JJA, the
sliding and jamming are {\it dual} in the sense that they are
simultaneously taking place along different axes. Indeed in \cite{vortex-jamming} it was found numerically that the shear-modulus is zero/finite along the direction of weaker/stronger coupling at $T=0$. Furthermore it was suggested that the symmetric point $\lambda=1$
is a critical point at zero temperature $\lambda_{c}(T=0)=1$ similar to
the jamming point in granular matters \cite{Nagel-group}.  
Indeed recent studies at finite temperatures suggest
$T_{c}(\lambda=1)=0$  \cite{Park-Choi-Kim-Jeon-Chung,Granato}.
On the other hand the growth of the peak height of the vortex structure
factor with the system size $L$ (Fig.~\ref{fig-structure-factor})
suggests $T_{c}(\lambda) > 0$ at least at $\lambda \gg 1$.  

The analogy to the FK model suggests that the analytic hull functions
$\phi_{x (y)}[y]$ becomes non-analytic at the critical point $\lambda=\lambda_{c}=1$
and remains non-analytic for $\lambda < 1$. (The other way around for
$\phi_{x (y)}[x]$). However we speculate that the $1/\lambda$
expansion, which yields only analytic hull functions, 
remains stable up to $\lambda=1$. Then certain ``non-perturbative''
solution(s) of \eq{eq-condition-plaquette} and \eq{eq-condition-vertex}
must emerge at weaker anisotropy and make level crossing(s)
with the horizontal stripe state.

To conclude we found undulated vortex stripes in irrationally frustrated
Josephson junction array with anisotropic Josephson coupling theoretically. 
It will be very interesting to study critical properties of the system closer to 
the symmetric point where the present perturbative approach should break down. 

\vspace*{.2cm}
{\bf Acknowledgement} We thank Hikaru Kawamura, Jorge Kurchan
and Hiroshi Matsukawa for useful discussions.
We thank the Supercomputer Center, ISSP, University of Tokyo for the use of the facilities.
This work is supported by Grant-in-Aid for Scientific Research
on Priority Areas "Novel States of Matter Induced by Frustration"
(1905200*) and Grant-in-Aid for Scientific Research (C) (21540386).


\begin{thebibliography}{99}
\bibitem{comment}
A system with e. g. 
$f=1/2+\epsilon$ with {\it infinitesimal irrational} $\epsilon$ should behave
as a system with $f=1/2$ in some sense. Although this is an interesting 
issue, we focus on  properties of systems
with irrational $f$ in $N\to \infty$ limit.
\end{thebibliography}
\end{document}